\pdfoutput=1

\documentclass[11pt]{article}

\usepackage{EMNLP2022}

\usepackage{times}
\usepackage{latexsym}
\usepackage{balance}
\usepackage{subfigure}
\usepackage{latexsym}
\usepackage{amsmath}
\usepackage{amssymb}
\usepackage{caption}
\usepackage{graphicx}
\usepackage{url}
\usepackage{hyperref}
\usepackage{multicol}
\usepackage{multirow}
\usepackage{booktabs}
\usepackage{color}
\usepackage{subfigure}
\usepackage{array}
\usepackage{mathtools} 
\usepackage[T1]{fontenc}

\usepackage[utf8]{inputenc}

\usepackage{microtype}

\usepackage{inconsolata}

\title{Dimension Reduction for Efficient Dense Retrieval via Conditional Autoencoder}

\author{Zhenghao Liu\textsuperscript{1}, 
        Han Zhang\textsuperscript{1}, 
        Chenyan Xiong\textsuperscript{2}, 
        Zhiyuan Liu\textsuperscript{3},
        Yu Gu\textsuperscript{1},
        \and Xiaohua Li\textsuperscript{1}\\
        $^1$Department of Computer Science and Technology, Northeastern University, China \\
        $^2$Microsoft Research, United States \\
        $^3$Department of Computer Science and Technology, Institute for AI, Tsinghua University, China \\
        Beijing National Research Center for Information Science and Technology, China \\
        \texttt{\{liuzhenghao,guyu,lixiaohua\}@mail.neu.edu.cn}\\
        \texttt{zhanghan@stumail.neu.edu.cn; chenyan.xiong@microsoft.com}\\
        \texttt{liuzy@tsinghua.edu.cn}}
        
\begin{document}

\maketitle
\begin{abstract}
Dense retrievers encode queries and documents and map them in an embedding space using pre-trained language models. These embeddings need to be high-dimensional to fit training signals and guarantee the retrieval effectiveness of dense retrievers. However, these high-dimensional embeddings lead to larger index storage and higher retrieval latency. To reduce the embedding dimensions of dense retrieval, this paper proposes a Conditional Autoencoder (ConAE) to compress the high-dimensional embeddings to maintain the same embedding distribution and better recover the ranking features. Our experiments show that ConAE is effective in compressing embeddings by achieving comparable ranking performance with its teacher model and making the retrieval system more efficient. Our further analyses show that ConAE can alleviate the redundancy of the embeddings of dense retrieval with only one linear layer. All codes of this work are available at \url{https://github.com/NEUIR/ConAE}.
\end{abstract}
\section{Introduction}
As the first stage of numerous multi-stage IR and NLP tasks~\cite{nogueira2019document,chen2017reading,thorne2018fact}, dense retrievers~\cite{xiong2020approximate} have shown lots of advances in conducting semantic searching and avoiding the vocabulary mismatch problem~\cite{robertson2009probabilistic}. Dense retrievers usually encode queries and documents as high-dimensional embeddings, which are necessary to guarantee retrieval effectiveness during training~\cite{ma2021simple,reimers2021curse}. Nevertheless, high dimensional embeddings usually exhaust the memory to store the index and lead to longer retrieval latency~\cite{indyk1998approximate,meiser1993point}.

The research of building efficient dense retrieval systems has been stimulated recently~\cite{effecientqa}. To reduce the dimensions of document embeddings, existing work reserves the principle dimensions or compresses query and document embeddings for building more efficient retrievers~\cite{yang2021designing,ma2021simple}.

There are two challenges in compressing embeddings of dense retrievers: The compressed embeddings should share a similar distribution with the original embeddings, making the low-dimensional embedding space uniform and the document embeddings distinguishable; All the compressed embeddings should have the ability to maintain the maximal information for matching related queries and documents during retrieval, which helps better align the related query-document pairs.

This paper proposes a Conditional Autoencoder (ConAE), which aims to build efficient dense retrieval systems by reducing the embedding dimensions of queries and documents. ConAE first encodes high-dimensional embeddings into a low-dimensional embedding space and then generates embeddings that can be aligned to related queries or documents in the original embedding space. In addition, ConAE designs a conditional loss to regulate the low-dimensional embedding space to mimic the embedding distribution of high-dimensional embeddings. Our experiments show that ConAE is effective to compress the high-dimensional embeddings and avoid redundant ranking features by achieving comparable retrieval performance with vanilla dense retrievers and better visualizing the embedding space with t-SNE.
\section{Related Work}
Dense retrievers use a bi-encoder architecture to encode queries and documents and map them in an embedding space for retrieval~\cite{karpukhin2020dense,xiong2020dense,xiong2020approximate,lewis2020pre,zhan2020learning,li2021more,Yu2021FewShotCD}. To learn an effective embedding space, dense retrievers are forced to maintain high-dimensional embeddings to fit training signals. 

The most direct way to reduce the dimension of embeddings is that retaining parts of the dimensions of high-dimensional embeddings~\cite{yang2021designing,ma2021simple}. Some work uses the first 128 dimensions to encode both queries and documents~\cite{yang2021designing} or utilizes PCA to retain the primary dimensions to recover most information from the raw embeddings~\cite{ma2021simple}. Other work~\cite{ma2021simple} proposes a supervised method, which uses neural networks to compress the high-dimensional embeddings as lower-dimensional ones. These supervised models provide a better dimension reduction way than unsupervised models by avoiding missing too much information. To optimize the encoders, some work~\cite{ma2021simple} continuously trains dense retrievers with the contrastive training strategies~\cite{karpukhin2020dense,xiong2020approximate}.
\section{Methodology}
This section introduces our Conditional Autoencoder (ConAE). We first introduce the preliminaries of dense retrieval (Sec.~\ref{model:preliminary}), and then describe the architecture of ConAE (Sec.~\ref{model:autoencoder}).

\subsection{Preliminary of Dense Retrieval}\label{model:preliminary}
Given a query $q$ and a document collection $D=\{d_1,\dots,d_j,\dots,d_n\}$, dense retrievers~\cite{xiong2020dense,xiong2020approximate,karpukhin2020dense} employ pre-trained language models~\cite{devlin2019bert,liu2019roberta} to encode $q$ and $d$ as $K$-dimensional embeddings, $h_q$ and $h_d$.

Then we can calculate the retrieval score $f(q,d)$ of $q$ and $d$ with dot product $f(h_q,h_d)=h_q \cdot h_d$. Then we contrastively train query and document encoders by maximizing the retrieval probability $P(d^+|q, \{d^+\} \cup D^-)$ of the relevant document $d^+$~\cite{xiong2020dense,xiong2020approximate}:
\begin{equation}\small
\label{eq:prob}
      P(d^+|q, \{d^+\} \cup D^-) = \frac{e^{f(h_q,h_{d^+})}}{e^{f(h_q,h_{d^+})} + \sum\limits_{d^-\in D^-}{e^{f(h_q,h_{d^-})}}},
\end{equation}
where $d^-$ is the document sampled from the irrelevant document set $D^-$~\cite{karpukhin2020dense,xiong2020approximate}.

\subsection{Dimension Compression with ConAE}\label{model:autoencoder}
In this subsection, we introduce ConAE to compress the $K$-dimensional embeddings $h_q$ and $h_d$ of both queries and documents to the $L$-dimensional embeddings $h_q^e$ and $h_d^e$.

\textbf{Encoder.} We first get the initial representations $h_q$ and $h_d$ for query $q$ and document $d$ from existing dense retrievers, such as ANCE~\cite{xiong2020approximate}. Then these $K$-dimensional embeddings can be compressed to low dimensional ones with two different linear layers, $\text{Linear}_q$ and $\text{Linear}_d$:
\begin{equation}
\small
     h_q^e = \text{Linear}_q(h_q); h_d^e = \text{Linear}_d(h_d).
\end{equation}
$h_q^e$ and $h_d^e$ are $L$-dimensional embeddings. The dimension $L$ can be 256, 128 or 64, which is much lower than the dimension $K$ of $h_q$ and $h_d$.  

Then we use KL divergence to regulate encoded embeddings to mimic the initial embedding distributions of queries and documents:
\begin{equation}\label{eq:kl}
\small
    L_{KL}  =  \sum\limits_{q}{\sum_{d \in D_{\text{top}}} P(d|q, D_{\text{top}}) \cdot \log  \frac{P(d|q, D_{\text{top}})}{P_e(d|q, D_{\text{top}} )}},
\end{equation}
where $P_e(d|q, D_{\text{top}})$ is calculated with E.q.~\ref{eq:prob}, using the encoded embeddings $h_q^e$ and $h_d^e$. $D_{\text{top}}$ consists of the top-ranked documents, which are searched by the teacher retriever--ANCE.
 
\textbf{Decoder.} The decoder module maps the encoded embeddings $h_q^e$ and $h_d^e$ into the original embedding space by aligning the compressed embeddings $h_q^e$ and $h_d^e$ with $h_q$ and $h_d$. It aims at optimizing encoder modules to maximally maintain ranking features from the initial representations $h_q$ and $h_d$ of query and document. 

Firstly, we use one linear layer to project $h_q^e$ and $h_d^e$ to $K$-dimensional embeddings, $\hat{h}_q$ and $\hat{h}_d$:
\begin{equation}\small
     \hat{h}_q = \text{Linear}(h_q^e);
     \hat{h}_d = \text{Linear}(h_d^e).
\end{equation}
Then we respectively train the decoded embeddings $\hat{h}_q$ and $\hat{h}_d$ to align with $h_q$ and $h_d$ in the original embedding space using two max margin losses $L_{q}$ and $L_{d}$. The max margin loss is widely used in previous neural IR research to optimize the ranking scores~\cite{xiong2017knrm,convknrm}.

The first loss $L_{q}$ is used to optimize the decoded query representation $\hat{h}_{q}$:
\begin{equation}\small
        L_{q}= \sum\limits_{q}{1 + \tanh f(\hat{h}_{q}, h_{d^-}) - \tanh f(\hat{h}_{q}, h_{d^+})},
\end{equation}
and we can also optimize the decoded document representation $\hat{h}_{d}$ with the second loss function $L_{d}$:
\begin{equation}\small
        L_{d}= \sum\limits_{q}{1 + \tanh  f(h_{q}, \hat{h}_{d^-}) - \tanh f(h_{q}, \hat{h}_{d^+})}.
\end{equation}

\textbf{Training Loss.} Finally, we train our conditional autoencoder model with the following loss $L$:
\begin{equation}
\small
        L= L_{KL} + \lambda L_{q} + \lambda L_{d},
\end{equation}
where $\lambda$ is a hyper-parameter to weight the autoencoder losses.
\begin{table}
\centering
\small
 \resizebox{0.48\textwidth}{!}{
\begin{tabular}{l  |  r r r r}
\hline  \multirow{2}{*}{{Dataset}} & \multirow{2}{*}{\#Doc} & \multicolumn{3}{c}{\#Queries} \\
& &{Train} & {Dev} & {Test}\\ \hline
MS MARCO & 8,841,823 & 452,939 & 50,000 & 6,980\\
NQ & 21,015,323 & 79,168 &8,757 & 3,610\\
TREC DL & 8,841,823 &- & - & 43 \\
TREC-COVID & 171,332 & - & - & 50\\
\hline
\end{tabular}}
\caption{Data Statistics.}
\label{tab:dataset}
\end{table}

\begin{table*}

\centering
\small
\resizebox{\linewidth}{!}{
\begin{tabular}{ l | c  c  c  c | c  c | c | c  }
\hline
\multirow{2}{*}{{Method}} & \multicolumn{4}{c|}{MS MARCO} & 	\multicolumn{2}{c|}{NQ}  &  {TREC DL} & {TREC-COVID} \\ 

& {Latency(ms)} & {MRR@10}  & {NDCG@10} & {Rec@1000} & {Top20} & {Top100}  & {NDCG@10} & {NDCG@10}\\ 
\hline
Teacher-768 & 17.152 & 0.3302 & 0.3877 & 0.9584 & 0.8224 & 0.8787 & 0.6489 & 0.6529\\ \hline
ANCE-256 & 6.159 & 0.3145 & 0.3709 & 0.9545 & \textbf{0.8188} & \textbf{0.8765} & \textbf{0.6455} & 0.5722 \\
PCA-256 & 6.296 & 0.2440 & 0.2940 & 0.9257 & 0.8042 & 0.8715 & 0.5118 & 0.2601 \\
CE-256 & 7.344 & 0.2959 & 0.3472 & 0.9333 & 0.8066 & 0.8726 & 0.5916 & 0.4110 \\
ConAE-256 & 7.158 & \textbf{0.3294}  & \textbf{0.3864} & \textbf{0.9560} & 0.8053 & 0.8723 & 0.6438 & \textbf{0.6405} \\ \hline
ANCE-128 & 3.419 & 0.3092 & 0.3667 & \textbf{0.9527} & \textbf{0.8069} &	\textbf{0.8709} &\textbf{0.6514} & 0.5612 \\
PCA-128 & 3.525 & 0.2348 & 0.2838 & 0.9170 & 0.7875 & 0.8620 &  0.4795 & 0.2523 \\
CE-128 & 4.530 & 0.2917 & 0.3438 & 0.9345 & 0.7934 & 0.8668 & 0.6170 & 0.4692 \\
ConAE-128& 3.942 & \textbf{0.3245} & \textbf{0.3816} & 0.9523 & 0.8064 & 0.8687 & 0.6380 & \textbf{0.6381} \\ \hline
ANCE-64 & 3.041 & 0.2773 & 0.3295 & 0.9217 & \textbf{0.7687} & \textbf{0.8474} & \textbf{0.6003} & 0.4731 \\
PCA-64 & 2.627 & 0.1855 & 0.2259 & 0.8540 & 0.6698 & 0.7928 & 0.3788 & 0.2174 \\
CE-64 & 3.046 & 0.2551 & 0.3036 & 0.9042 & 0.7404 & 0.8341 & 0.5561 & 0.3968 \\ 
ConAE-64& 3.087 & \textbf{0.2862} & \textbf{0.3376} & \textbf{0.9222} & 0.7604 & 0.8460 & 0.5877 & \textbf{0.5006} \\  \hline
\end{tabular}}
\caption{Performance of Different Dimension Reduction Models. We start from ANCE (Teacher), reduce the embedding dimension and evaluate their retrieval effectiveness. The document indices are built with flat index and the sizes of MS MARCO indices are 26.0G, 8.5G, 4.3G and 2.2G for 768, 256, 128 and 64 dimensional embeddings.}
\label{tab:overall}
\end{table*}

\section{Experimental Methodology}
This section describes the datasets, evaluation metrics, baselines and implementation details of our experiments.

\textbf{Dataset.} Four datasets are used to evaluate the retrieval effectiveness of different dimension reduction models, including MS MARCO (Passage Ranking)~\cite{bajaj2016ms}, NQ~\cite{kwiatkowski2019natural}, TREC DL~\cite{craswelloverview} and TREC-COVID~\cite{roberts2020trec}.
In our experiments, we randomly sample 50,000 queries from the raw training set of MS MARCO as the development set and use MS MARCO (Dev) as the testing set. The dimension reduction models that are trained on MS MARCO are also evaluated on two benchmarks, TREC DL and TREC-COVID, aiming to evaluate their generalization ability. All data statistics are shown in Table~\ref{tab:dataset}.

\textbf{Evaluation Metrics.} NDCG@10 is used as the evaluation metric on three benchmarks, MS MARCO, TREC DL and TREC-COVID. MS MARCO also uses MRR@10 as the primary evaluation metric~\cite{bajaj2016ms}. For the NQ dataset, the hit accuracy on Top20 and Top100 is used as the evaluation metric, which is the same as previous work~\cite{karpukhin2020dense}.

\textbf{Baselines.} In our experiments, we compare ConAE with two baselines from previous work~\cite{ma2021simple}, Principle Component Analysis (PCA) and CE. PCA reduces the embedding dimension by retaining the principle dimensions that can keep most of the variance within the original representation. CE model uses two linear layers $W_q$ and $W_d$ without biases to transform dense representations of queries and documents into lower embeddings~\cite{ma2021simple}. We also start from CE models and continuously train the whole model to implement our ANCE models to generate query and document embeddings of different dimensions.

\textbf{Implementation Details.}
The rest describes our implementation details. 
All embedding dimension reduction models base on one of the best dense retrievers ANCE~\cite{xiong2020approximate} and build document index with exact matching (flat index), which is implemented by FAISS~\cite{johnson2019billion}. During training ConAE, we set the hyper-parameter $\lambda$ as 0.1 and search Top100 documents using vanilla ANCE to construct the $D_{\text{top}}$ collection for each query. For our CE and ANCE models, we sample 7 negative documents for each query to contrastively train these models and sample 1 negative document to train ConAE. In our experiments, we set the batch size to 2 and accumulate step to 8 for ANCE. The batch size and accumulate step are 128 and 1 for other models. All models are implemented with PyTorch and tuned with Adam optimizer. The learning rates of ANCE and other models are set to $2e-6$ and 0.001, respectively.

\section{Evaluation Result}
Four experiments are conducted in this section to study the effectiveness of ConAE in reducing embedding dimensions for dense retrieval.

\begin{figure*}[t]
    \centering
    \subfigure[Teacher (t-SNE).] { \label{fig:tsne:768} 
    \includegraphics[width=0.23\linewidth]{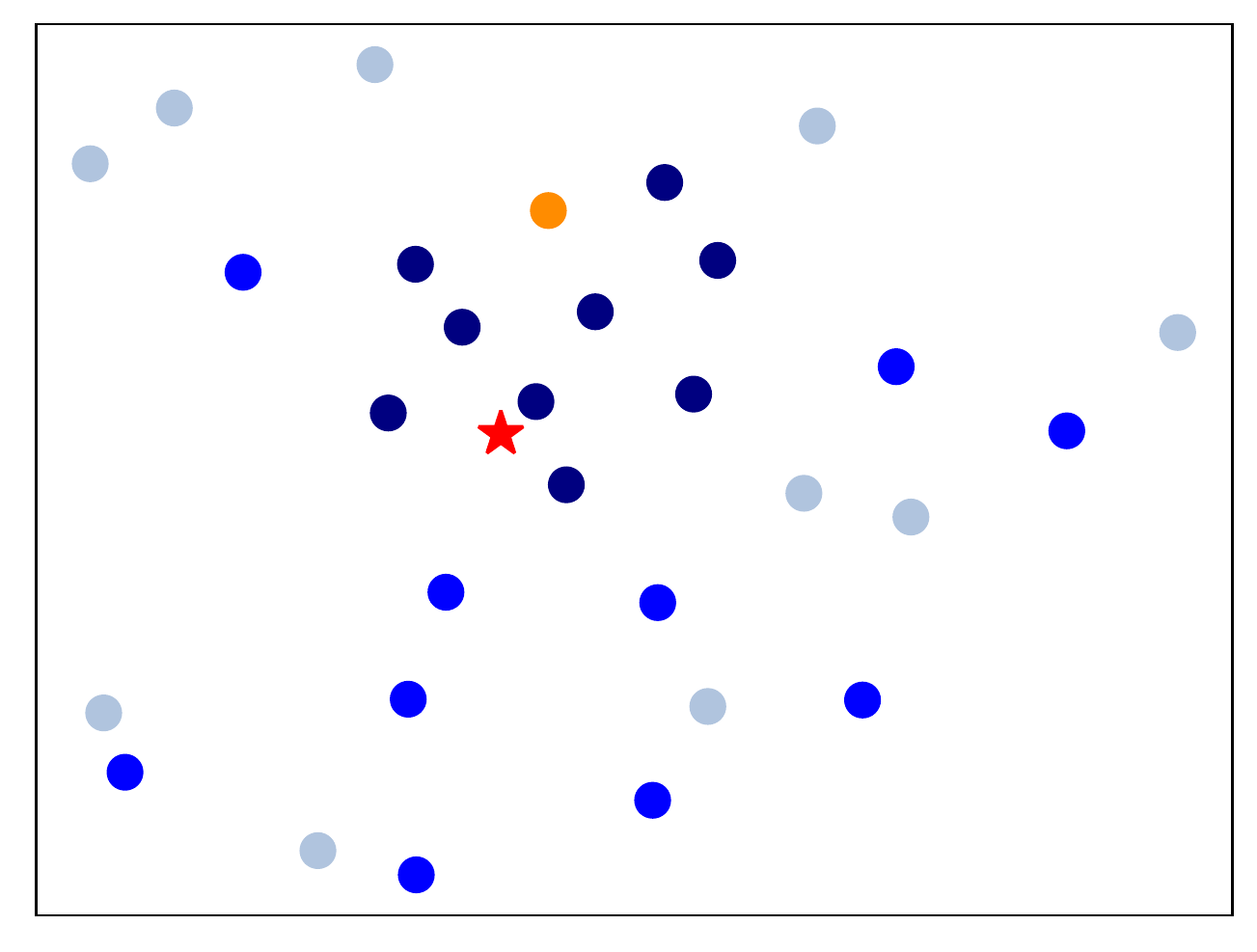}}
    \subfigure[ConAE-128 (t-SNE).] { \label{fig:tsne:128} 
    \includegraphics[width=0.23\linewidth]{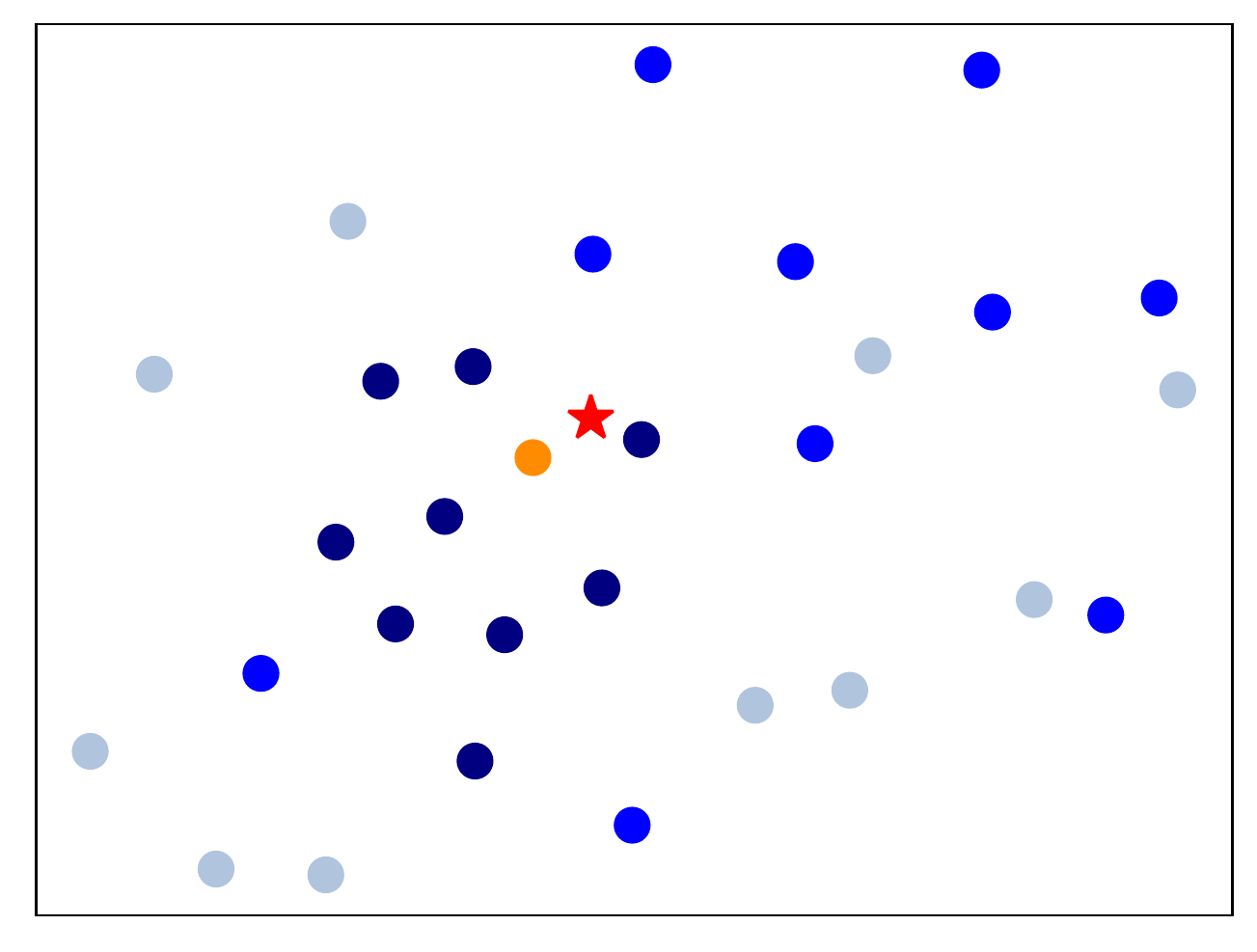}}
    \subfigure[ConAE-64 (t-SNE).] { \label{fig:tsne:64} 
    \includegraphics[width=0.23\linewidth]{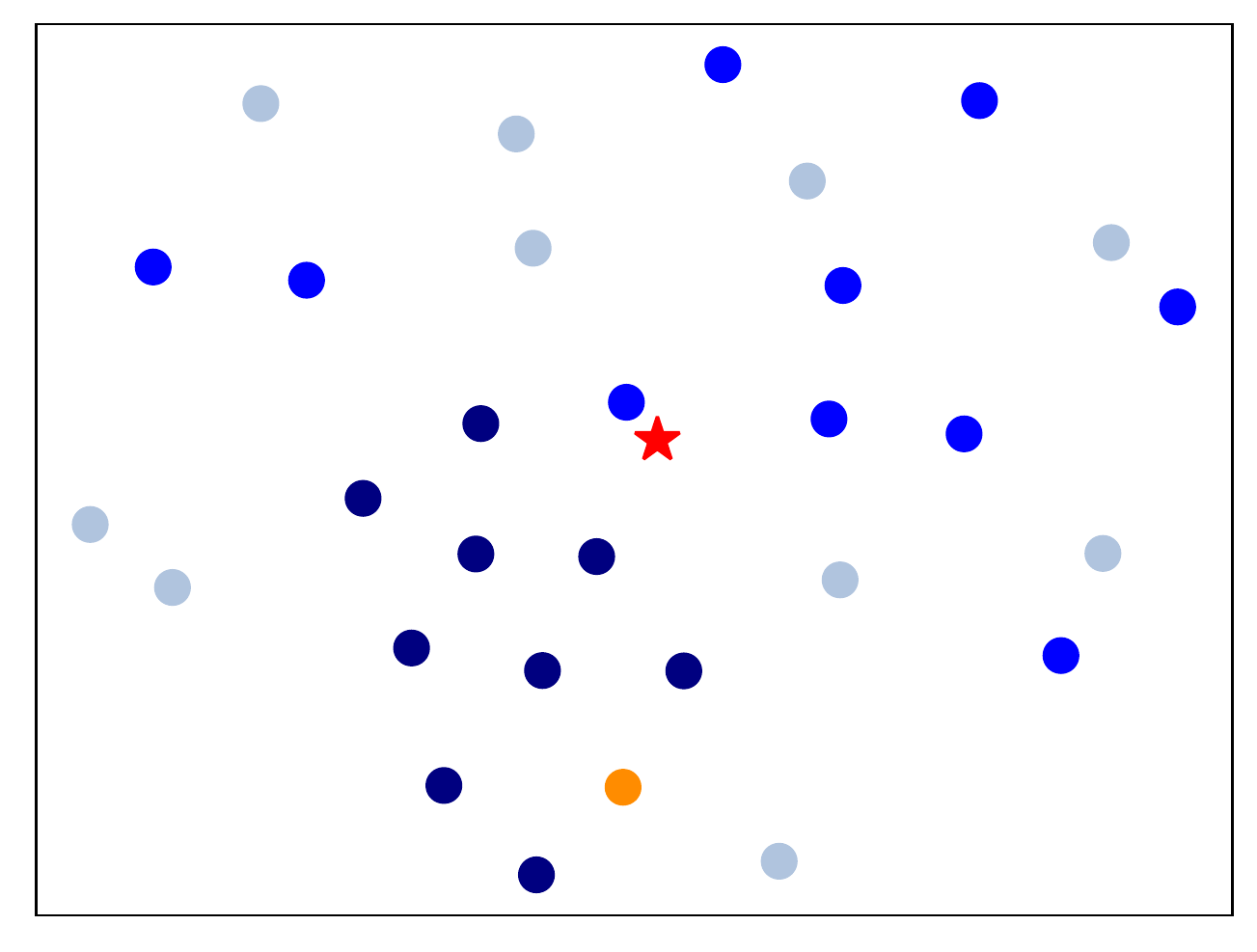}}
    \subfigure[Teacher (ConAE).] { \label{fig:tsne:2} 
    \includegraphics[width=0.23\linewidth]{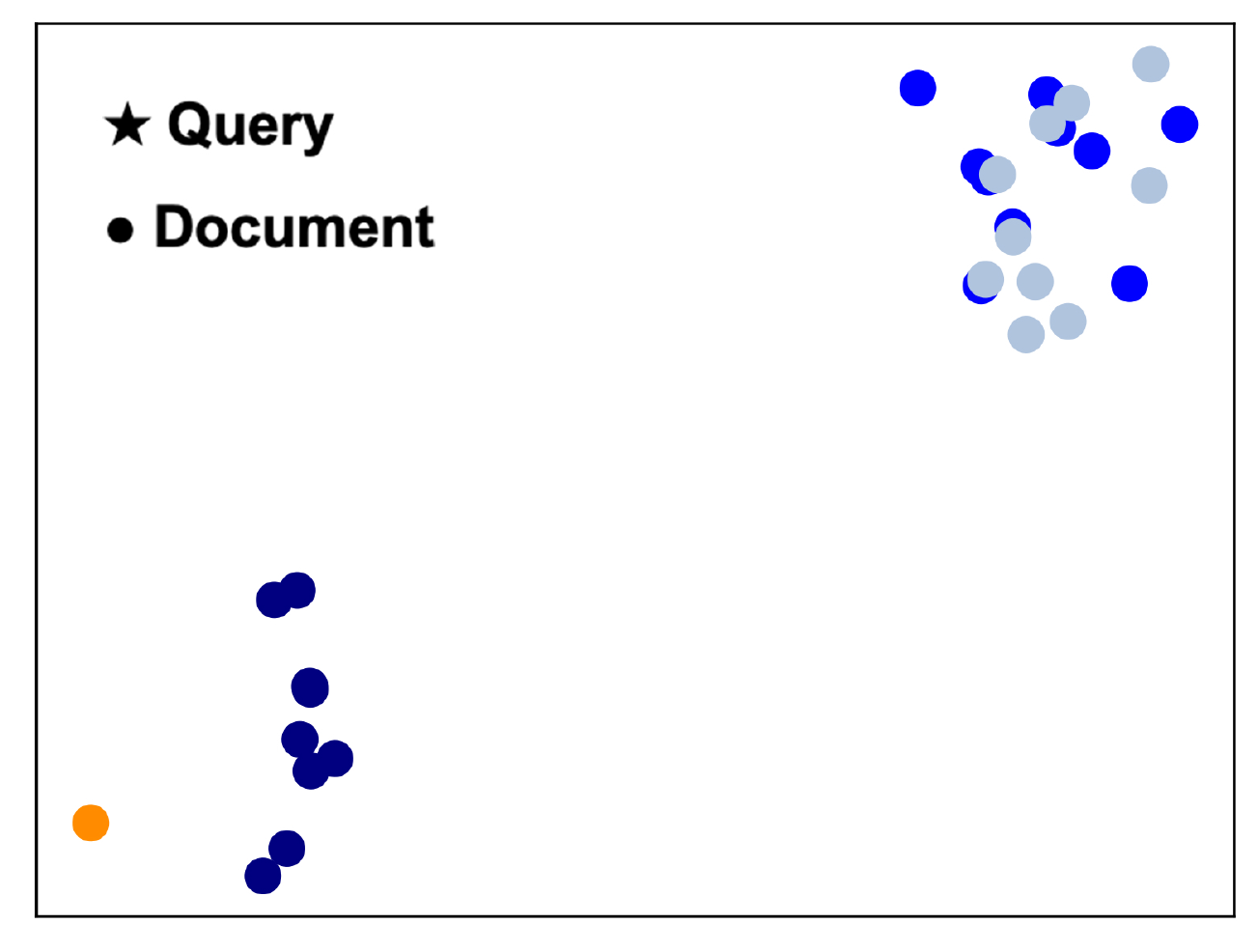}}
    \caption{Embedding Visualization of Different Dense Retrievers. Figure~\ref{fig:tsne:768}, \ref{fig:tsne:128} and \ref{fig:tsne:64} are plotted with t-SNE with 768, 128 and 64 dimensional embeddings. In Figure~\ref{fig:tsne:2}, we directly use ConAE w/o Decoder to visualize the document embedding space of ANCE. The ``$\bullet$'' in ``dark orange'' color denotes the golden document that ranked 2nd by ConAE-64 and 1st by other models. For other documents, darker blue ones are more relevant to the query.}
    \label{fig:tsne}
\end{figure*}

\subsection{Overall Performance}\label{sec:overall}
The performance of different dimension reduction models is shown in Table~\ref{tab:overall}. PCA, CE and ConAE are based on ANCE (Teacher), which freezes the teacher model and only optimizes the dimension projection layers. ANCE starts from CE and continuously tunes all parameters in the model.

Compared with PCA and CE~\cite{ma2021simple}, ConAE achieves the best performance on almost of datasets, which shows its effectiveness in compressing dense retrieval embeddings. ConAE can achieve comparable performance with ANCE (Teacher) using 128-dimensional embeddings to build the document index on MS MARCO, which reduces the retrieval latency (from 17.152 ms to 3.942 ms per query) and saves the index storage (from 26.0G to 4.3G) significantly. It demonstrates that ConAE is effective to alleviate the redundancy of the embeddings learned by dense retrievers.

Among all baselines, PCA shows significantly worse ranking performance on MS MARCO, indicating that embedding dimensions of dense retrievers are usually nonorthogonal. ConAE-128 achieves more than 11\% improvements than CE and performs much better on TREC-COVID, demonstrating its ranking effectiveness and generalization ability. ANCE can further improve the retrieval performance of CE by continuously training the query and document encoders, which adapts the teacher model to the low-dimensional version.

\subsection{Ablation Study}\label{sec:ablation}
This subsection conducts ablation studies in Table~\ref{tab:ablation} to investigate the effectiveness of different modules in our ConAE model.

\begin{table}
\centering
\small
\resizebox{\linewidth}{!}{
\begin{tabular}{ l | c  c  c }
\hline
\multirow{2}{*}{{Method}}& {MS MARCO} & 	{TREC-COVID}  &  {TREC DL}  \\ 
  & {MRR@10} & {NDCG@10}  & {NDCG@10} \\ 
\hline
ConAE-256 & \textbf{0.3294} & {0.6405} & 0.6438 \\
w/o Decoder & 0.3271 & \textbf{0.6546} & 0.6377  \\
w/o KL & 0.3276 & 0.6218 & \textbf{0.6491}  \\
\hline
ConAE-128 & \textbf{0.3245} & 0.6381 & \textbf{0.6380}  \\ 
w/o Decoder & 0.3203 & \textbf{0.6525} & 0.6266 \\
w/o KL & 0.3234 & 0.6365 & 0.6367 \\
\hline
ConAE-64 & \textbf{0.2862} & \textbf{0.5006} & 0.5877 \\ 
w/o Decoder & 0.2846 & 0.4703 & \textbf{0.5951} \\
w/o KL & 0.2822 & 0.4658 & 0.5759 \\
\hline
\end{tabular}}
\caption{Retrieval Performance of Different Ablation Models. ConAE w/o Decoder and ConAE w/o KL use $L_{KL}$ and $L_q + L_d$ to train the distillation models.}
\label{tab:ablation}
\end{table}

The different modules in ConAE play different roles. Compared with ConAE w/o Decoder, ConAE w/o KL usually shows better retrieval effectiveness on the two benchmarks MS MARCO and TREC DL, which ask the model to retrieve candidates from the same data source. It demonstrates that our autoencoder architecture can reserve more ranking features to fit the training supervision of MS MARCO. On the other hand, ConAE w/o Decoder shows stronger generalization ability by outperforming ConAE w/o KL on TREC-COVID, which belongs to a different domain. The source of the generalization ability of ConAE w/o Decoder may come from finer-grained training signals from our teacher model. The annotated training signals usually face the hole rate problem~\cite{xiong2020cmt} and using neural IR models to denoise the training signals has shown strong effectiveness in training neural IR models~\cite{DBLP:conf/naacl/QuDLLRZDWW21}.

ConAE combines both KL and autoencoder architectures to fully use training signals and regulate the distribution of compressed embedding, which usually achieves better retrieval performance.



\subsection{Embedding Visualization with ConAE}
We randomly sample one case from MS MARCO and visualize the embedding space of query and retrieved documents in Figure~\ref{fig:tsne}.

We first employ t-SNE~\cite{van2008visualizing} to visualize the embedding spaces of ANCE (Teacher) and ConAE. As shown in Figure~\ref{fig:tsne:128}, ConAE-128 conducts a more meaningful visualization results: the related query-document pair is closer and the other documents are distributed around the golden document according to their relevance to the query. The visualization of ANCE (Teacher) is slightly distorted and different from our expectations, which is mainly due to its redundancy. The redundant features usually mislead t-SNE to overfit these ranking features, thus reducing the embedding dimension of dense retrievers to 128 provides a possible way to alleviate redundant features and better visualize the embedding space of dense retrievers using t-SNE. Besides, ConAE-64 shows decreased retrieval performance than ConAE-128 (Sec.~\ref{sec:overall}).
As shown in Figure~\ref{fig:tsne:64}, it mainly derives from that ConAE-64 loses some ranking features with the limited embedding dimensions. 

The other way to visualize the embedding space is using ConAE w/o Decoder to project the embedding to a 2-dimensional coordinate. It uses KL divergence to optimize the 2-dimensional embeddings to mimic the relevance score distribution of teacher models. As shown in Figure~\ref{fig:tsne:2}, the distributions of documents are distinguishable, which provides an intuitive way to analyze the ranking-oriented document distribution. In addition, the query is usually far away from the documents. The main reason lies that the relevance scores are calculated by dot product and the embedding norms are meaningful to distinguish the relevant documents.

\subsection{Retrieval Performance with HNSW}\label{app:ann}
\begin{table}
\small
\centering
\begin{tabular}{ c | l | c | c  c  }
\hline 
\multirow{2}{*}{{Dim.}} & \multirow{2}{*}{{Method}} & {Latency} & \multicolumn{2}{c}{{MS MARCO}} \\

& & {(ms)} &{MRR@10} & {NDCG@10} \\ 
\hline
768 &ANCE &2.056 & 0.3295 & 0.3869\\
\hline
\multirow{3}{*}{256} & PCA & \textbf{1.016} & 0.2427 & 0.2924 \\
& CE & 1.646 & 0.2948 & 0.3461 \\
& ConAE & 1.478 & \textbf{0.3257} & \textbf{0.3818} \\ \hline
\multirow{3}{*}{128} & PCA & \textbf{0.702} & 0.2340 & 0.2829\\
& CE & 1.047 & 0.2906 & 0.3424 \\
& ConAE & 0.978 & \textbf{0.3209} & \textbf{0.3770} \\ \hline
\multirow{3}{*}{64} & PCA & \textbf{0.616} & 0.1844 & 0.2246 \\
& CE & 0.860 & 0.2545 & 0.3030 \\
& ConAE & 0.983 & \textbf{0.2837} & \textbf{0.3344} \\ \hline

\end{tabular}
\caption{ANN Retrieval Effectiveness of Different Models. The ANN index is built with HNSW.}
\label{tab:hnsw}
\end{table}
Besides exact searching, we also show retrieval results of different dimension reduction methods in Table~\ref{tab:hnsw}, which are implemented by the approximate nearest neighbor (ANN) search, Hierarchical Navigable Small World (HNSW). Using HNSW, the retrieval efficiency can be further improved, especially for high-dimensional embeddings. ConAE keeps its advanced retrieval performance again with less than 1ms retrieval latency.
\section{Conclusion}
This paper presents ConAE, which reduces the embedding dimension of dense retrievers. Our experiments show that ConAE can achieve comparable retrieval performance with the teacher model, significantly reduce the index storage and accelerate the searching process. Our further analyses show that the high-dimensional embeddings of dense retrievers are usually redundant and ConAE helps to alleviate such redundancy and visualize the embedding space more intuitively and effectively.
\section*{Limitations}
In this paper, we mainly focus on compressing the embeddings of dense retrievers in an additional stage between query/document encoding and index building. As a result, we fix query and document embeddings of dense retrievers and project high-dimensional embeddings to low-dimensional ones using only one linear layer. Thus, the effectiveness of ConAE is limited by the number of learnable parameters. Even though ConAE shows comparable performance with ANCE (Teacher), joint modeling the query/document encoder, dimension reduction module and index building still show strong potential to achieve better retrieval performance.

\section*{Acknowledgments}
This work is mainly supported by Beijing Academy of Artificial Intelligence (BAAI) as well as supported in part by the Natural Science Foundation of China under Grant No. 62206042 and No. 62006129, the Fundamental Research Funds for the Central Universities under Grant N2216013, China Postdoctoral Science Foundation under Grant 2022M710022 and National Science and Technology Major Project (J2019-IV-0002-0069).

\bibliography{custom}
\bibliographystyle{acl_natbib}

\end{document}